\documentclass[runningheads]{svmult}

\usepackage{makeidx}   
\usepackage{graphicx}  
\usepackage{subeqnar}  
\usepackage{multicol}  
\usepackage{physprbb}  
\makeindex             

\begin{document}
\title*{Chandra Observations of the Components of Clusters, Groups, and
Galaxies and their Interactions\thanks{Contribution for the
{\em Lighthouses of the Universe Conference}, 6-10 August 2001,
Garching bei Munchen, Germany}}
\titlerunning{Chandra Observations of Clusters, Groups, and
Galaxies}
%
\author{W. Forman\inst{1}
\and C. Jones\inst{1}
\and M. Markevitch\inst{1,3}
\and A. Vikhlinin\inst{1,3}
\and E. Churazov\inst{2,3}
}
\authorrunning{W. Forman et al.}
%
\institute{Smithsonian Astrophysical Observatory, 60 Garden St. Cambridge, MA USA
\and Max-Planck-Institut f\"ur Astrophysik, Karl-Schwarzschild-Strasse
1, 85741 Garching, Germany
\and Space Research Institute (IKI), Profsoyuznaya 84/32, Moscow 117810, Russia
}

\maketitle              

\begin{abstract}
We discuss two themes from Chandra observations of galaxies, groups,
and clusters. First, we review the merging process as seen through the
high angular resolution of Chandra. We present examples of sharp,
edge-like surface brightness structures ``cold fronts'', the
boundaries of the remaining cores of merger components and the Chandra
observations of CL0657, the first clear example of a strong cluster
merger shock. In addition to reviewing already published work, we
present observations of the cold front around the elliptical galaxy
NGC1404 which is infalling into the Fornax cluster and we discuss
multiple ``edges'' in ZW3146. Second, we review the effects of
relativistic, radio-emitting plasmas or ``bubbles'', inflated by
active galactic nuclei, on the hot X-ray emitting gaseous atmospheres
in galaxies and clusters. We review published work and also discuss
the unusual X-ray structures surrounding the galaxies NGC4636 and
NGC507.
\end{abstract}

\section{Introduction}

With its first images, the Einstein Observatory changed our view of
clusters and galaxies.  Clusters, rather than being dynamically old,
relaxed systems showed extensive substructure reflecting complex
gravitational potentials with ``double'' clusters merging on times
scales of $\sim10^9$~yrs~\cite{jones1979,jones1984,wrf1}.
Luminous elliptical galaxies, rather than being gas poor, were instead found to
be gas rich systems with {\em hot} coronae having masses up to
$\sim10^9$ M$_\odot$~\cite{wrf2}. ROSAT and ASCA continued the
revolution with studies of cluster merging and substructure including
Coma, A2256, A754, Cygnus A, Centaurus, A1367,and Virgo, to mention
just a
few~\cite{briel1991,briel1992,bohringer1994,alexey2,henry1995,honda1996,henriksen1996,alexey1,chur1999,mark1999,donnelly,schindler1999,makishima2001}.

The high angular resolution provided by Chandra has again brought us
new views of old friends -- early type galaxies and clusters of
galaxies -- that we thought we knew pretty well. We are familiar with
the ingredients, galaxies, radio emitting plasma, hot gas, and dark
matter. The recipe is simple -- {\em mix vigorously}. With these
simple instructions we find new and unexpected phenomena in the
Chandra observations.

\section{A New Aspect of Cluster Mergers -- Cold Fronts}

For many years clusters were thought to be dynamically relaxed systems
evolving slowly after an initial, short-lived episode of violent
relaxation. However, in a prescient paper, Gunn \&
Gott argued that, while the dynamical timescale for the
Coma cluster, the prototype of a relaxed cluster, was comfortably less
than the Hubble time, other less dense clusters had  dynamical
timescales comparable to or longer than the age of the Universe~\cite{gunn1972}. Gunn
\& Gott concluded that ``The present is the epoch of cluster
formation''.  With the launch of the Einstein Observatory came the
ability to ``image'' the gravitational potential around clusters. Many
papers in the 1980's exploited the imaging capability of the Einstein
Observatory and showed the rich and complex structure of galaxy
clusters. 

The X-ray observations supported the now prevalent idea that structure
in the Universe has grown through gravitational amplification of small
scale instabilities or hierarchical clustering. At one extreme, some
clusters grow, in their final phase, through mergers of nearly equal
mass components. Such mergers can be spectacular events involving
kinetic energies as large as $\sim 10^{64}$ ergs, the most energetic
events since the Big Bang.  More common are smaller mergers and
accretion of material from large scale filaments.  An example showing
the relationship between large scale structure and cluster merging is
seen in the ROSAT image of A85 where small groups are detected,
infalling along a filament into the main cluster.  The central cluster
galaxy, the bright cluster galaxies, an X-ray filament and nearby
groups and clusters all show a common alignment at a position angle of
about $160^{\circ}$ extending from 100 kpc (the outer isophotes of the
central cD galaxy) to 25 Mpc (the alignment of nearby
clusters)~\cite{durret1998}. Such common alignments over a wide range
of scales are expected if clusters form through accretion of matter
from filaments~\cite{haarlem1993}.

Chandra's high angular resolution has further illuminated the merging
process and the complexity of the X-ray emitting intracluster medium
(ICM).  Prior to the launch of Chandra, sharp gas density
discontinuities had been observed in the ROSAT images of
A2142 and A3667~\cite{mark1999}. Since both clusters
exhibited characteristics of major mergers, these features were
expected to be shock fronts. However, the first Chandra observations
showed that these were not shocks, but a new kind of structure -- cold
fronts~\cite{mark2000}. Their study has provided new and detailed
insights into the physics of the ICM~\cite{vik2001a,vik2001b}.

\subsection{Cold Fronts in Cluster Mergers}

The first cold front observed by Chandra was in the hot ($kT \sim 9$
keV), X-ray-luminous cluster A2142 ($z=0.089$). Two bright elliptical
galaxies, whose velocities differ by $1840$ km s$^{-1}$ lie near the cluster
center and are aligned in the general direction of the X-ray
brightness elongation~\cite{mark2000}. A2142 exhibits two sharp surface
brightness edges -- one lies $\sim 3'$ northwest of the cluster center
(seen earlier in the ROSAT image) and a second lies $\sim 1'$ south of
the center~\cite{mark2000}. The gas temperature distributions across
the edges show sharp and significant {\em increases} as the surface
brightness (gas density) {\em decreases}~\cite{mark2000}. The gas
density changes across the edges compensate, within the uncertainties,
for the temperature increases so that the gas pressures across the
edges are consistent with being equal.  One possible origin of the
A2142 structures is that they arise from the merger of two
inter-penetrating systems whose dense cores have survived the merger
process~\cite{mark2000}. We observe A2142 as it would appear after the
shock fronts have passed by each of the dense cores. The outer, lower
density gas has been shock heated, but the dense cores remain
``cold''~\cite{mark2000}. Each sharp edge is then a boundary of a ram
pressure-stripped subcluster core.

\begin{figure}
\centerline{\includegraphics[width=0.5\textwidth,bb=90 185 500 610,clip]{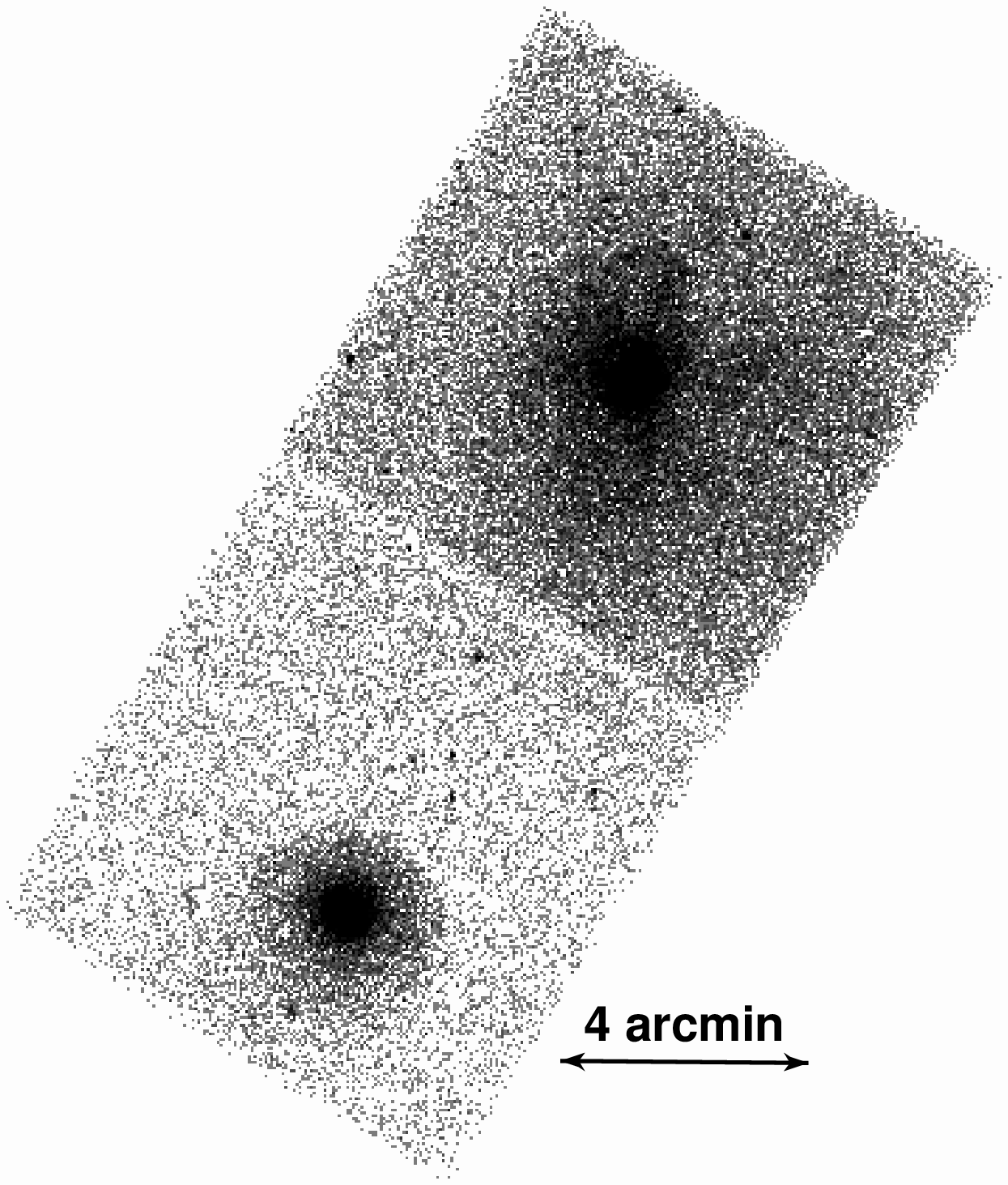}
\includegraphics[width=0.5\textwidth,bb=111 200 464 549,clip]{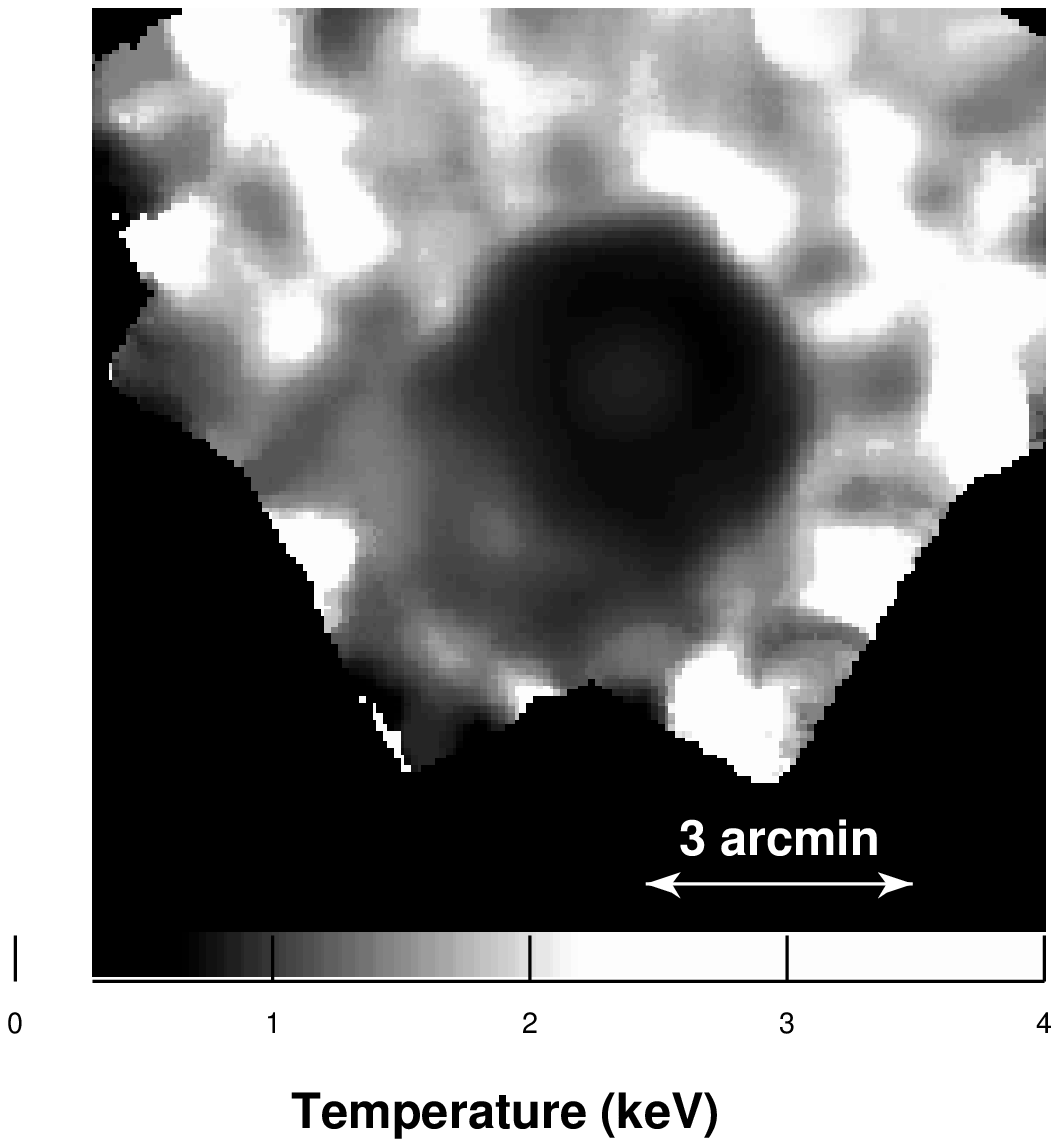}}
\caption{The ACIS observation of NGC1404 and NGC1399. (\textbf{a}) 
  shows the 0.5--2.0 keV band image of the Fornax cluster.  The gas
  filled dark halo surrounding NGC1404 is at the lower left
  (southeast) while the cluster core, dominated by the halo surrounding
  NGC1399 lies at the upper right (northwest). (\textbf{b}) The
  temperature map of the Fornax region. The cold core surrounding
  NGC1404 has a temperature of less $\sim1$ keV while the
  surrounding gas has a temperature of $_>\atop^{\sim}$1.5 keV.  }
\label{n1404_image}
\label{n1404_tmap}
\end{figure}

A particularly beautiful example of a cold front is seen in the
Chandra observation of the Fornax cluster.  In Fig.~\ref{n1404_image}a,
we see gas bound to the infalling bright elliptical galaxy NGC1404 as it
approaches the cluster center (to the northwest). The image clearly
shows the sharp edge of the surface brightness discontinuity, shaped
by the ram pressure of the cluster gas.  The temperature map
(Fig.~\ref{n1404_tmap}b) confirms that the infalling cloud is cold
compared to the hotter Fornax ICM.

Another example of a galaxy infalling into a cluster potential is M86.
X-ray emission from M86 was first observed with the Einstein
Observatory and subsequently by ROSAT. Its unusual ``plume'' was
explained as a ram pressure stripped galactic corona produced as M86
crosses the core of the Virgo cluster at supersonic
velocity~\cite{wrf1979,fabian1980,ros1,ros2}. The Chandra image and
temperature map confirm a cool ($\sim0.6$ keV) corona embedded in the
hot ($\sim3$ keV) Virgo ICM~\cite{moriond}. Based on the difference
between the velocity of M86 and the Virgo  cluster core, about
$\sim$1500~km~sec$^{-1}$, M86 must be crossing the Virgo core at supersonic
velocity.  However, we see no clear cold front as in NGC1404 (for the
image and temperature map of M86 see Fig.~6 and Fig.~7 in Forman
et al.~\cite{moriond}).  The most likely explanation is that M86 is
moving nearly in the line of sight, directly towards us, and the
resulting cold front (and possible shocks) are difficult to see in
projection.

\subsection{Cluster Physics and Edges}

\begin{figure} 
  
  \centerline{\includegraphics[width=0.85\textwidth]{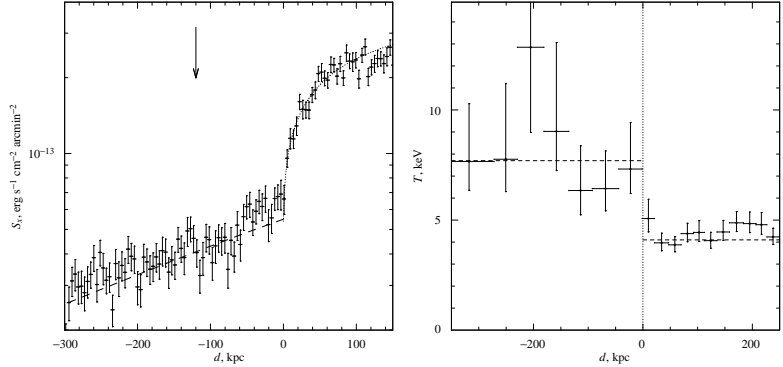}}
  \caption{(\textbf{a}) The surface brightness profile of A3667
  extracted in elliptical regions across the cold front. The 
  sharp ``edge'' is clearly seen. The dashed line is the ROSAT PSPC
  fit to the outer surface brightness distribution and agrees well
  with the Chandra observation. The dotted curve is a fit to a
  spheroid with a sharp boundary. The vertical arrow indicates the
  position of the weak shock. As discussed by Vikhlinin et al., the
  excess at distances of 0-50 kpc in front of the edge represents gas
  that accumulates in the stagnation region~\cite{vik2001a}.
  (\textbf{b}) The temperature profile across the cold front. The
  temperature {\em increases} from $\sim 4$ keV to $\sim 8$ keV across
  the front.  }
\label{a3667_profs}
\end{figure}

A3667, a moderately distant cluster ($z=0.055$),
also was expected to exhibit a shock front based on its ROSAT
image~\cite{mark1999}. However, as with A2142, the sharp feature is the
boundary of a dense cold cloud, a merger
remnant~\cite{vik2001a,vik2001b}. As Vikhlinin et al.  showed, the edge
is accurately modeled as a spheroid (see
Fig.~\ref{a3667_profs}a)~\cite{vik2001a}. From the surface brightness
profile, converted to gas density, and precise gas temperatures, the
gas pressure on both sides of the cold front can be accurately
calculated. The difference between the two pressures is a measure of
the ram pressure of the ICM on the moving cold front.  Hence, the
precise measurement of the gas parameters yields the cloud velocity.
The factor of two difference in pressures between the free streaming
region and the region immediately inside the cold
front yields a Mach number of the cloud of $1\pm0.2$ ($1430\pm290$ km
s$^{-1}$)~\cite{vik2001a}.

In addition to the edge, a weak shock is detected.  The distance
between the cold front and the weak shock ($\sim$350 kpc) and the
observed gas density jump at the shock (a factor of 1.1-1.2) yield the
shock's propagation velocity, $\sim$1600 km s$^{-1}$, which is consistent with
that derived independently from the pressure jump across the cold
front~\cite{vik2001a}.

The A3667 observation provides important information on the
efficiency of transport processes in clusters. As the 
surface brightness profile shows (see Fig.~\ref{a3667_profs}), the density ``edge''
is very sharp. Quantitatively, Vikhlinin et al. found that the width
of the front was less than $3.5''$ (5 kpc). This sharp edge requires
that transport processes across the edge be suppressed,
presumably by magnetic fields. Without such suppression, the edge
should be broader since the relevant Coulomb mean free path for electrons
is about 13 kpc, several times the width of the cold front~\cite{vik2001a}.
Furthermore, Vikhlinin et al. observed that the cold front appears
sharp only over a sector of about $\pm30^{\circ}$ centered on the
direction of motion, while at larger angles, the sharp boundary
disappears~\cite{vik2001b}. The disappearance can be explained by the
onset of Kelvin-Helmholtz instabilities, as the ambient ICM gas flows past
the moving cold front. To explain the observed extent of the sharp
boundary, the instability must be partially suppressed, e.g., by a
magnetic field parallel to the boundary with a strength of
$7-16\mu$G. Such a parallel magnetic field may be drawn out by the
flow along the front. This measured value of the magnetic field in the
cold front implies that the pressure from magnetic fields is small
(only 10-20\% of the thermal pressure) and, hence, supports
the accuracy of cluster gravitating mass estimates derived from X-ray
measurements that assume that the X-ray emitting gas is in hydrostatic
equilibrium and is supported by thermal pressure~\cite{vik2001b}.

\begin{figure}
\centerline{\includegraphics[width=0.450\textwidth,bb= 111 270 477
520,clip]{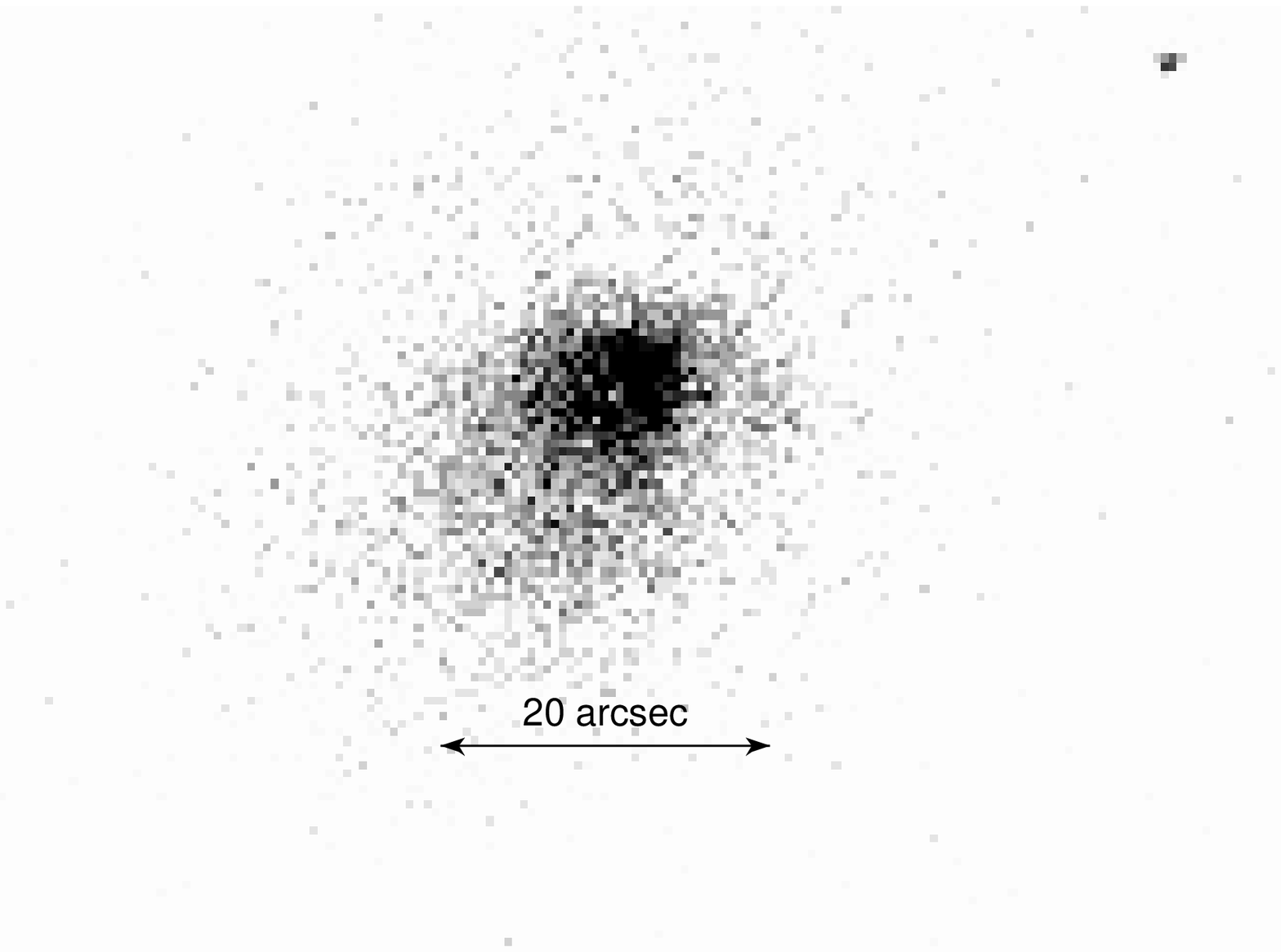}\includegraphics[width=0.450\textwidth,bb= 63
220 500 500,clip]{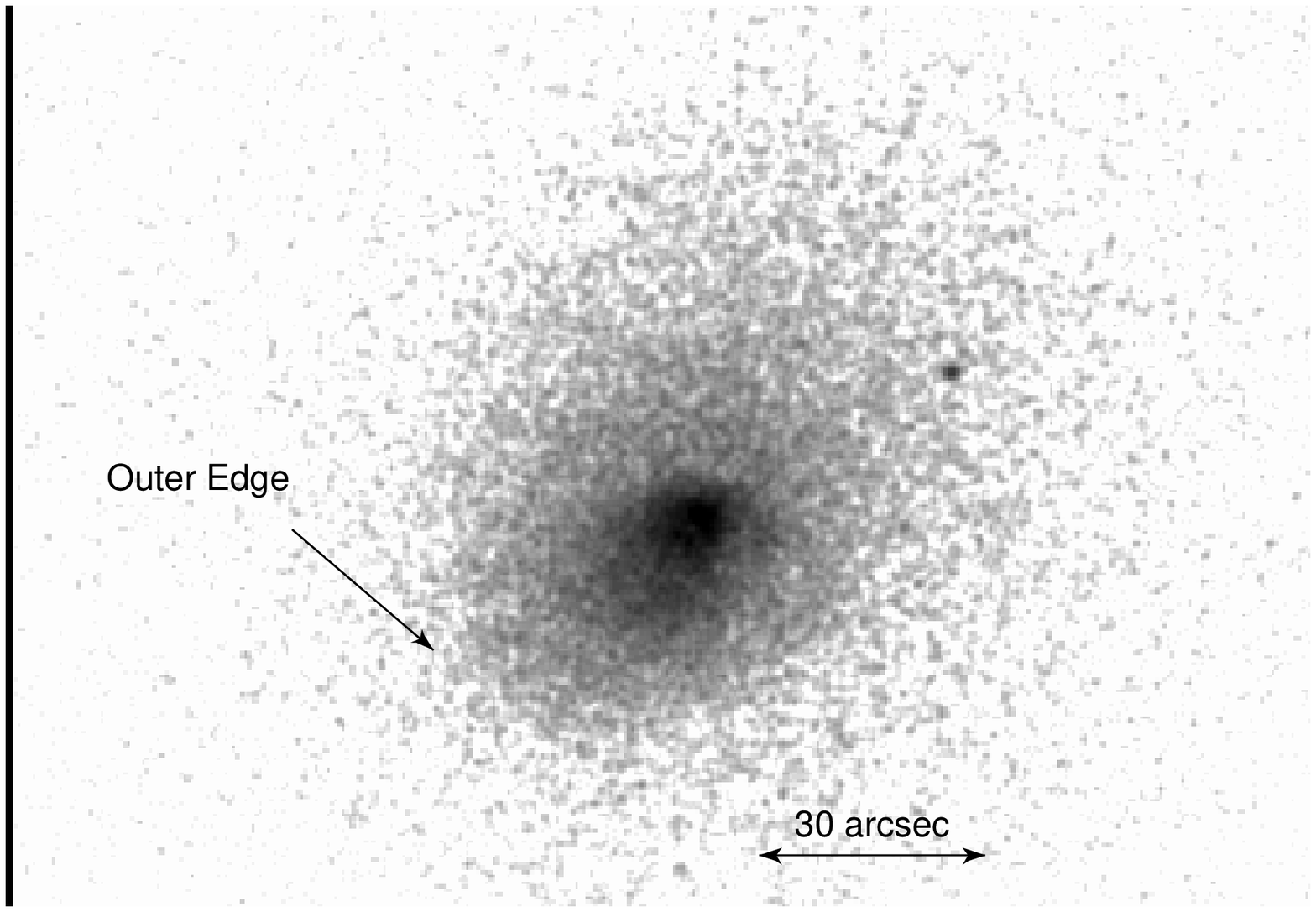}}
\caption{Multiple X-ray surface brightness edges in ZW3146 ($z=0.296$). 
  (\textbf{a}) The 0.5-2.0 keV image of the central
  region of ZW3146 shows the two inner edges at $3''$ and $8''$.
  (\textbf{b}) The right panel shows the edge at $35''$. }
\label{zw3146}
\end{figure}

\subsection{ZW3146 -- Multiple Cold Fronts}

ZW3146 is a moderately distant ($z=0.2906$; 5.74 kpc per arcsec)
cluster with a remarkably high mass deposition rate that was estimated
to exceed 1000 M$_{\odot}$~yr$^{-1}$~\cite{edge}. The Chandra image further
demonstrates the remarkable nature of this cluster -- on scales from
$3''$ to $30''$ ($\sim20$ kpc to 170 kpc), three separate X-ray
surface brightness edges are
detected (see Fig.~\ref{zw3146} and Forman et al.~\cite{moriond}). At the
smallest radii, two edges are seen to the northwest and north of the
center (see
Fig.\ref{zw3146}a).  The first, at a radius of $\sim3''$
(17 kpc), spans an angle of nearly $180^{\circ}$ with a surface brightness
drop of almost a factor of 2.  The second edge, at a radius of
$\sim8''$ (45 kpc) spans only $90^{\circ}$ but has a surface brightness
drop of almost a factor of 4.  The third edge (see 
Fig.~\ref{zw3146}b) lies to the southeast, about $35''$ (200 kpc) from
the cluster center, has a decrease of about a factor of 2, and, as
with the first edge, extends over an angle of almost $180^{\circ}$.

The variety of morphologies and scales exhibited by these sharp edges
or cold fronts is quite remarkable. Possibly the edges may arise from
moving cold gas clouds that are the remnants of merger activity as
observed in A2142 and A3667 or as oscillations (or ``sloshing'') of
the cool gas at the center of the cluster potential as observed in
A1795~\cite{a1795}. The extremely regular morphology of ZW3146 on
large linear scales seems to exclude a recent merger and, hence,
``sloshing'' of the gas seems the more likely explanation for the
observed edges.  High resolution, large scale structure simulations
show that dense halos, formed at very early epochs, would not be
disrupted as clusters collapse~\cite{ghigna1,ghigna2}. While most of
the dark matter halos, having galaxy size masses, are associated with
the sites of galaxy formation, the larger mass halos also may survive
(without their gas) or may have fallen into the cluster only
recently. Hence, we might expect to find a range of halo mass
distributions moving within the cluster potential. We speculate that,
as these halos move, the varying gravitational potential could accelerate
the cool dense gas that has accumulated in the cluster core and could
produce the ``sloshing'' needed to give rise to the multiple surface
brightness edges observed in some clusters. Simulations are needed to
confirm such a possibility.

\subsection{CL0657 -- A Prototypical Cluster Shock Front}

CL0657 ($z=0.296$) was discovered by Tucker et al. as part of a search
for ``failed'' clusters, clusters that were X-ray bright but had
few, if any, optical galaxies~\cite{tucker1}. From ASCA observations,
this cluster was found to have a remarkably hot gas temperature of
about 17 keV, making it the hottest cluster known~\cite{tucker2}.

\begin{figure} [ht]
\centerline{\includegraphics[width=0.50\textwidth]{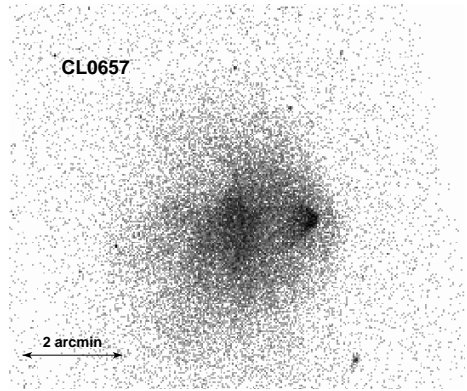}}
\caption{The Chandra image of the cluster CL0657.  The cluster
  exhibits the classic properties of a supersonic merger -- a dense
  (cold) ``bullet'' traversing the hot cluster with a leading shock front (Mach
  cone). The gas parameters across the front imply the cold core is
  traversing the cluster at a supersonic velocity with a Mach number
  of $M\sim2-3$. The disrupted core of the cluster can be seen to the
  east of the ``bullet''.
}
\label{0657}
\end{figure}

The Chandra image of CL0657 shows the classic properties of a
supersonic merger (see Markevitch et al. for a detailed discussion of
this cluster~\cite{maxim0657}).  We see a dense (cold) core moving to
the west after having traversed, and disrupted, the core of the main
cluster. Leading the cold, dense core is a density discontinuity which
appears as a shock front (Mach cone) and is confirmed by the spectral
data to be hotter to the east (trailing the shock), unlike the cold
fronts discussed above (or the eastern boundary of the bullet which
also is a cold front). The detailed gas density parameters confirm that
the ``bullet'' is moving to the west with a velocity of
3000--4000~km~sec$^{-1}$, 
approximately 2-3 times the sound speed of the ambient
gas. CL0657 is the first clear example of a relatively strong shock
arising from cluster mergers.

The Chandra observation confirms the unusually high mean cluster
temperature of 14-15 keV, but shows regions with temperatures as high
as 20 keV. The unrelaxed nature of CL0657 urges caution in the use of
high temperatures of a few extreme clusters to derive cosmological
constraints. Since the present mean temperature is likely to differ
from what it will become after the cluster achieves hydrostatic
equilibrium, the derived cluster mass would be in error and would
result in  incorrect cosmological constraints.

\section{The Radio---X-ray Connection -- or Bubbles, Bubbles Everywhere}

Prior to the launch of Chandra, ROSAT observations of NGC1275 and M87
provided hints of complex interactions between radio emitting plasmas
ejected from AGN within the nuclei of dominant, central cluster
galaxies~\cite{boh1993,boh1995,chur2000a}. With the launch of Chandra,
the interaction between the radio emitting plasma and the hot
intracluster medium (ICM) has been observed in many systems and now can
be studied in detail. 

\subsection{Hot Plasma Bubbles in Cluster and Galaxy Atmospheres}

One of the first, and clearest, examples of the effect of plasma
bubbles on the hot intracluster medium was found in the Perseus
cluster around the bright active, central galaxy NGC1275 (3C84). First
studied in ROSAT images\cite{boh1993}, the radio emitting cavities to
the north and south of NGC1275 are clearly seen in the Chandra images
with bright X-ray emitting rims surrounding the cavities that coincide
with the inner radio lobes~\cite{fabian2000}. For NGC1275/Perseus, the
radio lobes are in approximate pressure equilibrium with the ambient,
denser and cooler gas and the bright X-ray rims surrounding the
cavities are softer than the ambient gas.  The central galaxy in the
Hydra A cluster also harbors X-ray cavities associated with radio
lobes that also show no evidence for shock heating~\cite{mcnamara2000}.
Both sets of radio bubbles, being of lower density than the ambient
gas, must be buoyant.

The Chandra images of Perseus/NGC1275 also suggest the presence of
older bubbles produced by earlier outbursts~\cite{fabian2000}. These
older bubbles appear as X-ray surface brightness ``holes'', but unlike
the inner bubbles, these outer holes show no detectable radio
emission, suggesting that the synchrotron emitting electrons may have
decayed away leaving a heated, plasma bubble (see Fabian et al. who
recently reported low frequency radio spurs extending towards the
outer bubbles in NGC1275, consistent with this
scenario~\cite{bubbles}).  Such bubbles, with no attendant radio
emission, are seen by Chandra in the galaxy groups HCG62 and
MKW3s~\cite{vrtilek2001,mazzotta2001}.

The examples of bubbles described above concentrate on those around
central dominant cluster galaxies. However, bubbles, and their effects
are seen in more common early type galaxies.  For example, in the E1
galaxy M84 (NGC4374), Chandra observed an unusual X-ray morphology
which is explained by the effect of the radio lobes on the hot
gas~\cite{finoguenov2001}. The X-ray emission appears ${\cal
  H}$-shaped, with a bar extending east-west with two nearly parallel
filaments perpendicular to this bar.  The complex X-ray surface
brightness distribution arises from the presence of two radio lobes
(approximately north and south of the galaxy) that produce two low
density regions surrounded by higher density X-ray filaments. As with
Perseus/NGC1275 and Hydra A, the filaments, defining the ${\cal
  H}$-shaped emission, have gas temperatures comparable to the gas in
the central and outer regions of the galaxy and hence argue against
any strong shock heating of the galaxy atmosphere by the radio plasma.
By deriving the gas density surrounding the radio lobes, Finoguenov
\& Jones were able to calculate the strength of the magnetic field using the
observed Faraday rotation. They inferred a line-of-sight magnetic
field of $0.8 \mu$ Gauss~\cite{finoguenov2001}.

\subsection{Evolution of Buoyant Plasma Bubbles in Hot Gaseous Atmospheres}

The 327 MHz high resolution, high dynamic range radio map of M87 shows
a well-defined torus-like eastern bubble and a less well-defined
western bubble, both of which are connected to the central emission by
a column, and two very faint almost circular emission regions
northeast and southwest of the center~\cite{owen2000}. The correlation
between X--ray and radio emitting features has been remarked by several
authors~\cite{feigelson1987,boh1995,harris1999}.

Motivated by the similarity in appearance between M87 and hot bubbles
rising in a gaseous atmosphere, Churazov et al. developed a simple
model of the M87 bubbles which is generally applicable to the many
bubble-like systems seen in the Chandra observations~\cite{chur2001}.
An initial buoyant, spherical bubble transforms into a torus as it
rises through the galaxy or cluster atmosphere.  By entraining cool
gas as it rises, it exhibits a characteristic ``mushroom'' appearance,
similar to an atmospheric nuclear explosion.  This may qualitatively
explain the correlation of the radio and X--ray emitting plasmas and
naturally accounts for the thermal nature of the X-ray emission
associated with the rising torus~\cite{boh1995}.
Finally, in the last evolutionary phase of an atmospheric explosion,
the bubble reaches a height at which the ambient gas density equals
that of the bubble.  The bubble then expands to form a thin layer (a
``pancake'').  The large low surface brightness features in the M87 radio
map could be just such pancakes -- the final evolutionary phase of the
bubbles. In the simulations performed by 
Churazov et al. the 
buoyant bubbles behaved  as expected and did produce the features
observed in both X-rays and radio for M87. Although the exact form of
the rising bubbles was sensitive to initial conditions, the toroidal
structures were a common feature. Ambient gas was uplifted in the
cluster atmosphere reducing the effects of gas cooling and flowing to
the center and producing the ``stem'' of the mushroom that is
brighter than the surrounding regions~\cite{chur2001}.

\subsection{Explosive Cavities}

NGC4636 is one of the nearest and most X-ray luminous ``normal''
elliptical galaxies ($L_X \sim$2$\times 10^{41}$ ergs s$^{-1}$).  The first
X-ray imaging observations of NGC4636 from Einstein showed that, like
other luminous elliptical galaxies, NGC4636 was surrounded by an
extensive hot gas corona~\cite{wrf2}.

\begin{figure}
\centerline{\includegraphics[width=0.50\textwidth]{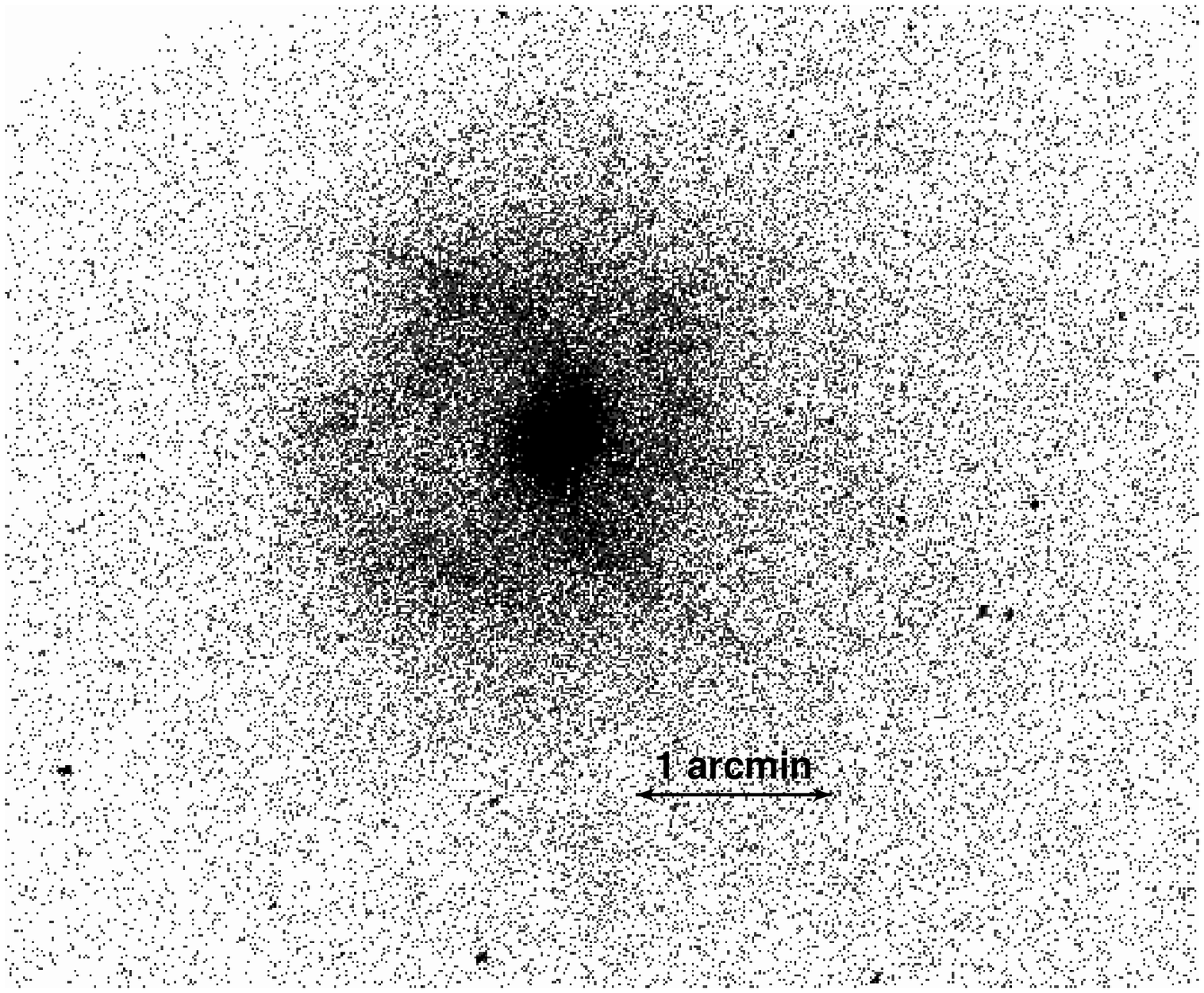}
\includegraphics[width=0.50\textwidth]{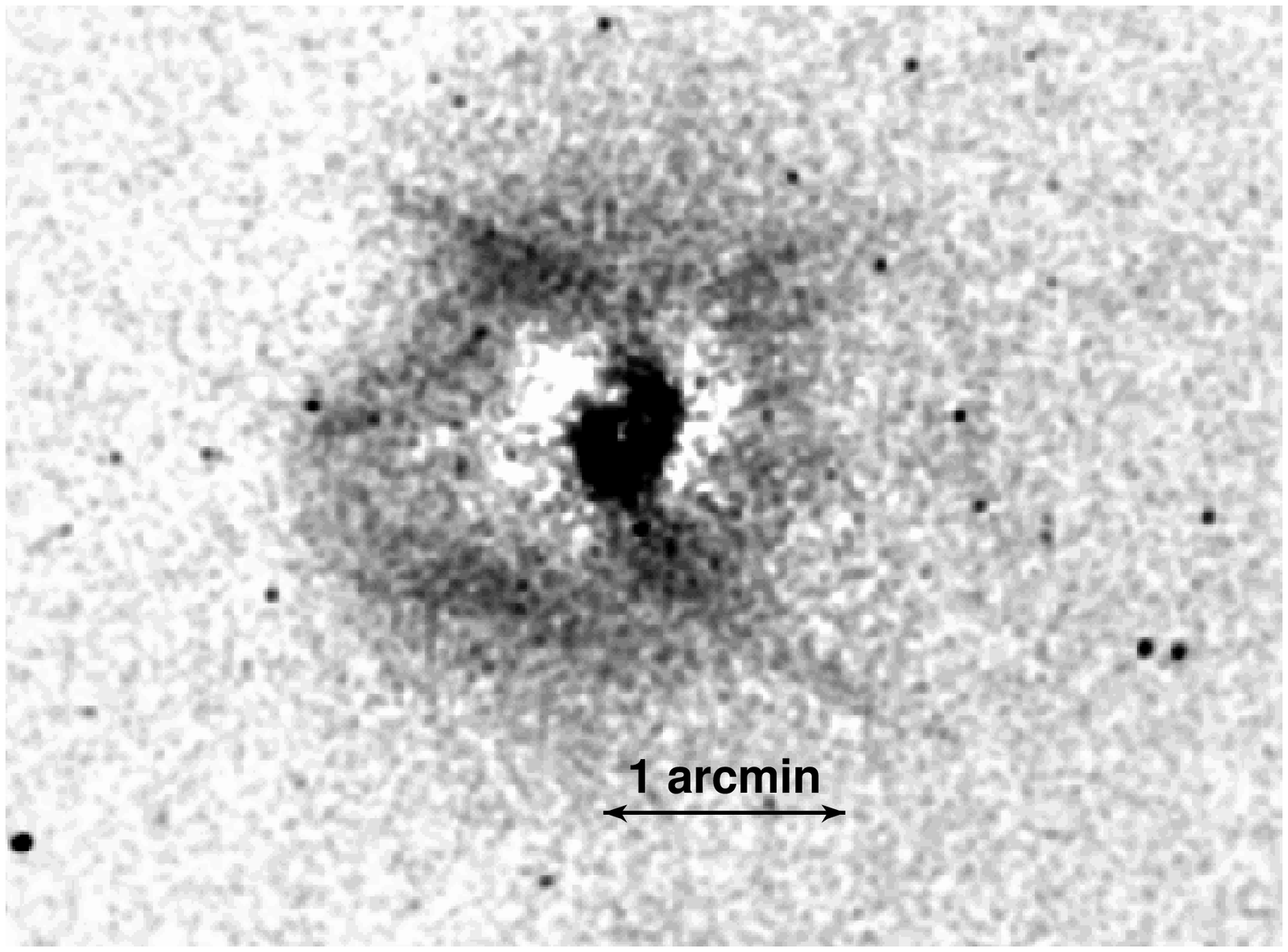}}
\caption{(\textbf{a}) shows the 0.5-2.0 keV ACIS-S image of NGC4636 
  at full resolution (1 pixel = $0.492''$). (\textbf{b}) shows the
  emission after an azimuthally symmetric model describing the galaxy
  corona has been subtracted.  The remaining emission was smoothed
  with a two pixel Gaussian. Shocks from a nuclear outburst
  could produce the brighter arm-like structures, while the additional features
  could arise from other outbursts.  }
\label{n4636_image}
\label{n4636_subrat}
\end{figure}

The Chandra observation of NGC4636 shows a new phenomenon -- shocks
produced by nuclear outbursts (see Jones et al. for details of the
Chandra observation of NGC4636~\cite{n4636}). The high angular resolution
Chandra image (see Fig.~\ref{n4636_image}) shows symmetric, $\sim8$
kpc long, arm-like features in the X-ray halo surrounding NGC4636.
The leading edges of these features are sharp and are accompanied by
temperature increases of $\sim30$\%, as expected from shocks
propagating in a galaxy atmosphere.

Although the sharpness of the edges of the NE and SW arms appears
similar to the sharp edges found along ``fronts'' in clusters (see
discussion and references above), the cluster ``fronts'' are cold,
while those in NGC4636 are hot.  Also, while the presence of sharp
fronts suggests the possibility of an ongoing merger, the east-west
symmetry of the halo structures, the similarity of this structure to
that seen around radio lobes, as well as the lack of a disturbed
morphology in the stellar core or in the stellar velocities suggest an
outburst from the nucleus as the underlying cause.  In particular, the bright
SW arm, the fainter NW arm and the bright NE arm can be produced by
the projected edges of two paraboloidal shock fronts expanding about an
east -- west axis through the nucleus.  A shock model is also
consistent with the evacuated cavities to the east and west of the
central region.

The size, symmetry, and gas density and temperature profiles of the
shocks are consistent with a nuclear outburt of energy $\sim 6 \times
10^{56}$ ergs having occurred about $\sim 3\times10^6$ years ago.  It is
tempting to suggest that these outbursts are part of a cycle in which
cooling gas fuels nuclear outbursts that periodically reheat the
cooling gas. Such outbursts if sufficiently frequent could prevent the
accumulation of significant amounts of cooled gas in the galaxy center.

\begin{figure}
\centerline{\includegraphics[width=0.3\textwidth,bb=75 220 450
  550,clip]{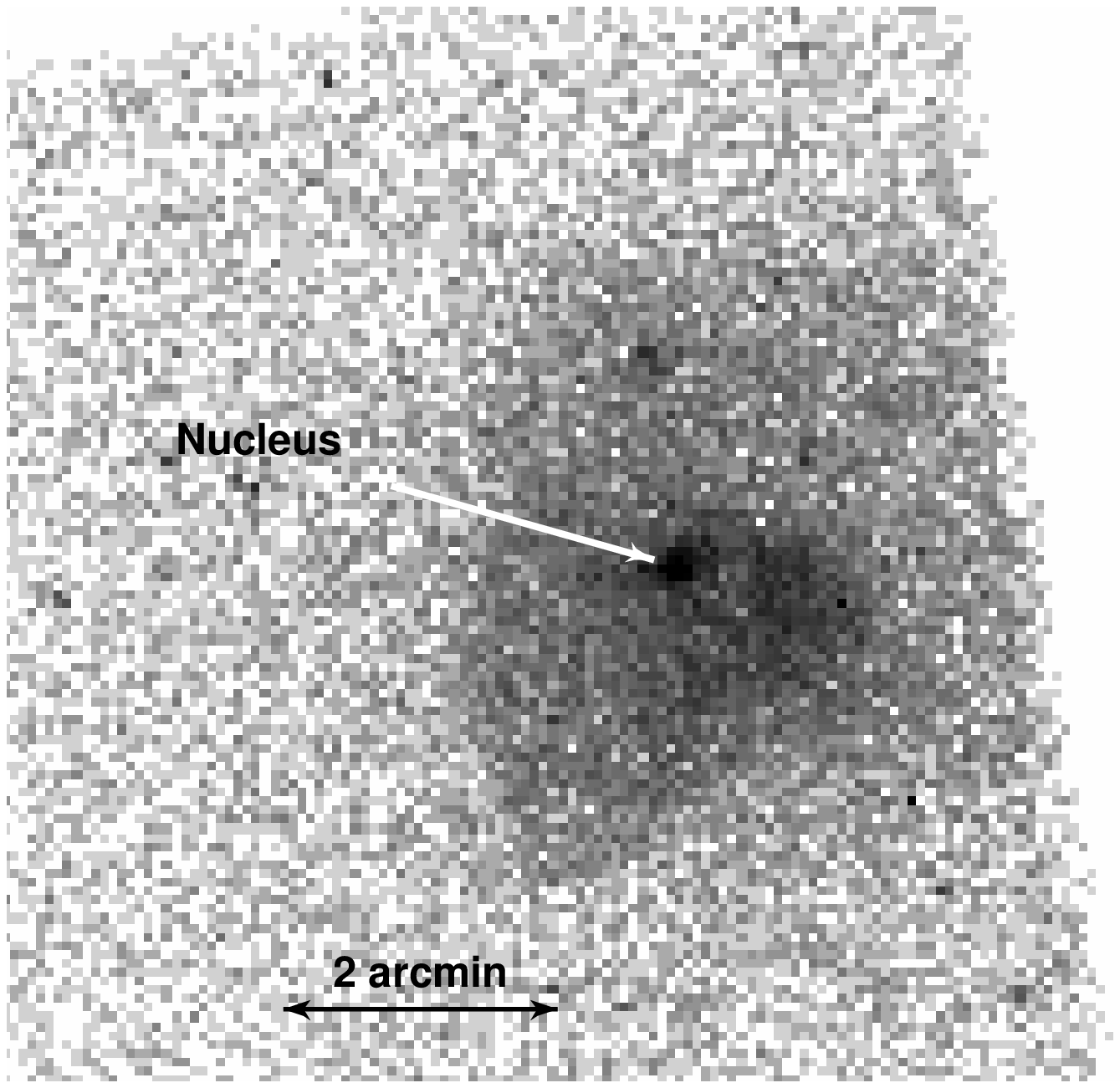}
\includegraphics[width=0.3\textwidth,bb=79 175 440 500,clip]{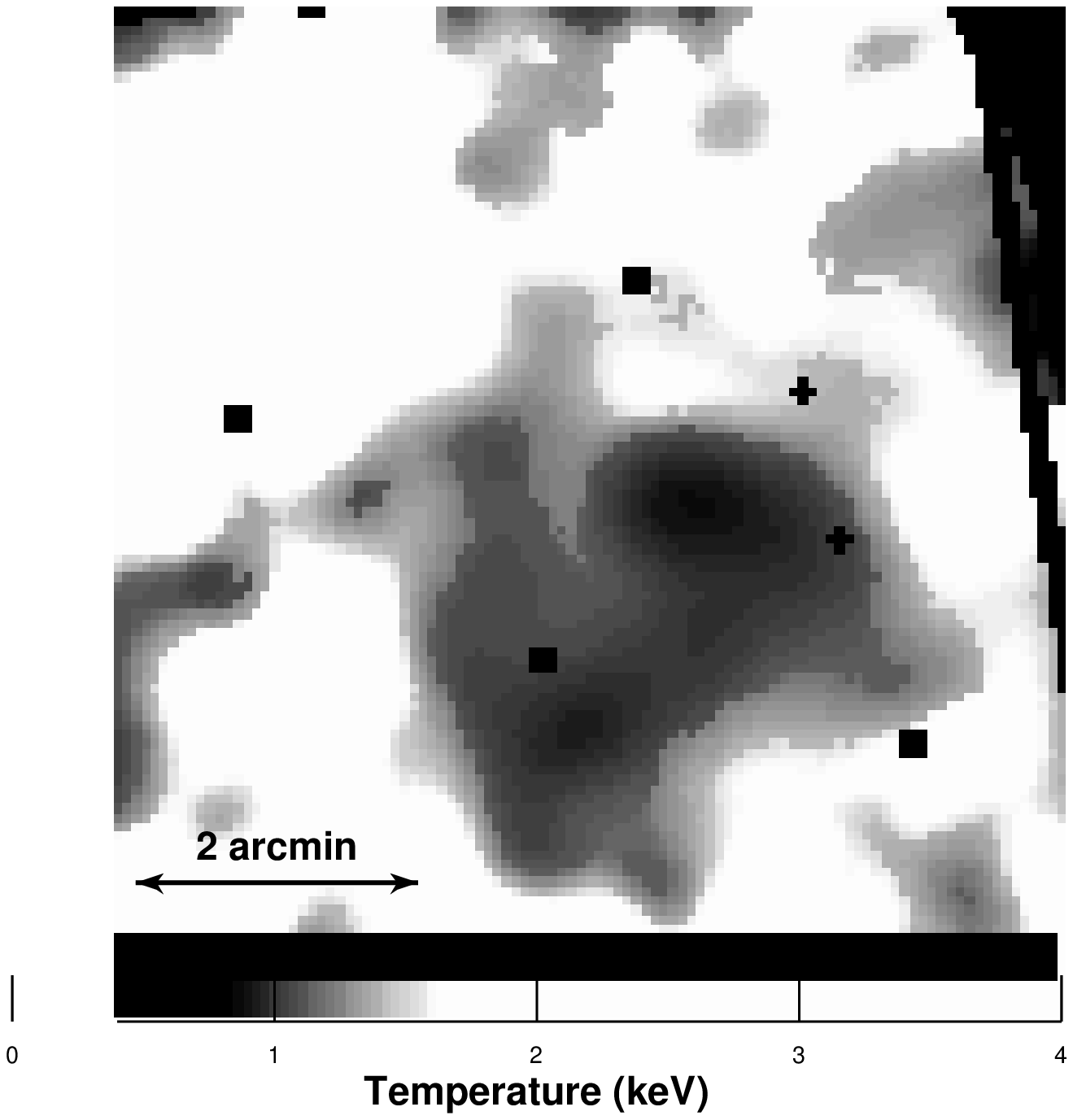}
\includegraphics[width=0.3\textwidth,bb=23 66 573 528,clip]{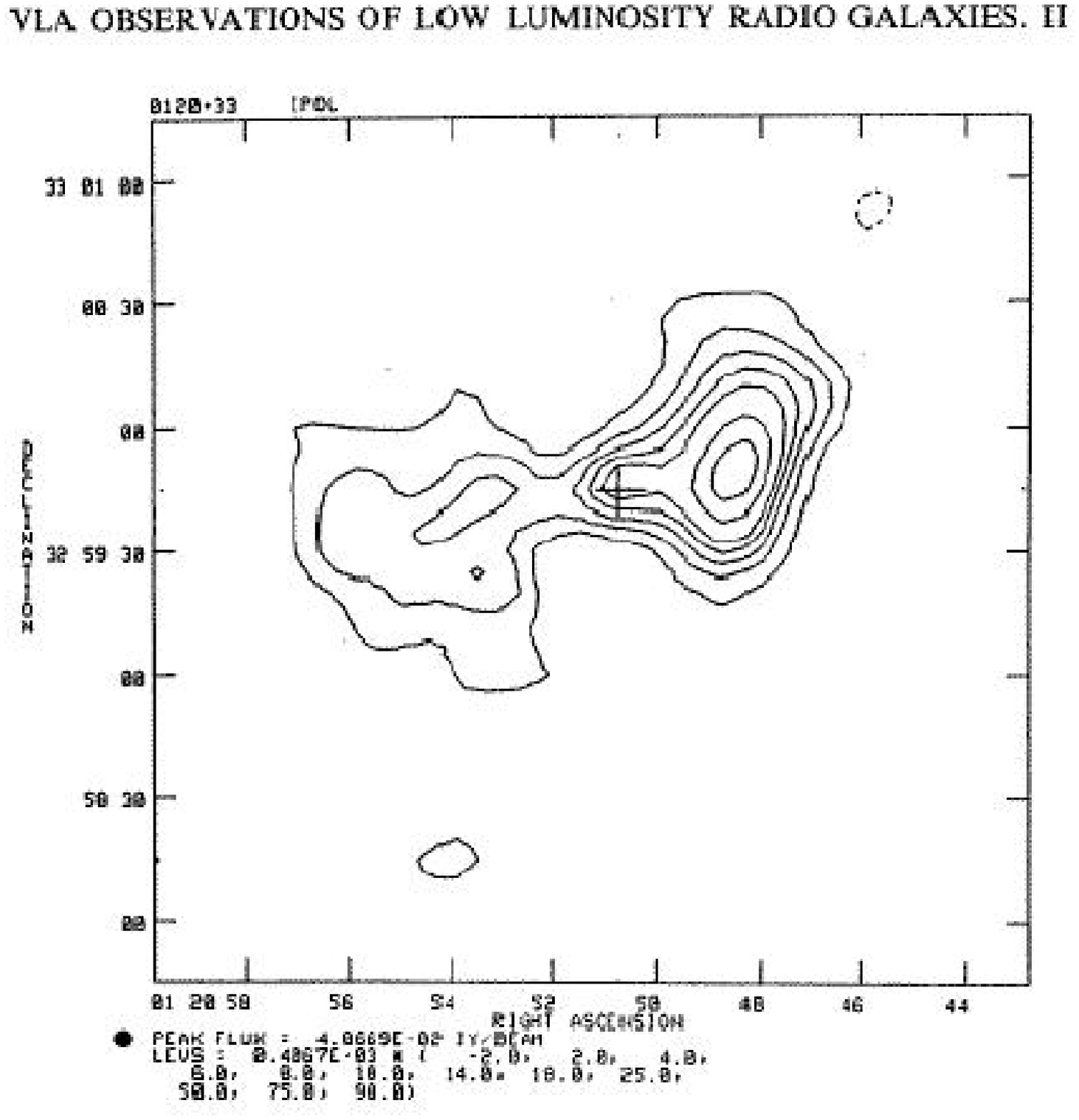}}
\caption{(\textbf{a}) The 0.5-2.0 keV surface brightness distribution of
  NGC507. (\textbf{b})  The temperature map of the central
  region of NGC507.  The galaxy center is cool as are the region to
  the west, the north-south edge to the east, and the edge to the
  south running from northwest to southeast. (\textbf{c}) The VLA
  radio map showing the central point source, a jet emanating to the
  west and two radio lobes~\cite{ruiter}.  The
  depression in the X-ray surface brightness to the west of the galaxy
  peak coincides with the western radio lobe. 
  }
\label{n507}
\end{figure}

\subsection{NGC507 - the central galaxy in a group}

NGC507 is the central galaxy in a nearby ($z=0.016$) group
that has been studied extensively in
X-rays~\cite{kim1995,matsumoto1997,buote1998,fukazawa1998}. The galaxy
is the site of a weak B2 radio source (luminosity $\sim10^{37}$ ergs
s$^{-1}$)~\cite{ruiter}. The Chandra X-ray image, shown in
Fig.~\ref{n507}a, covers only the central, high surface
brightness emission of the group around NGC507.  The 0.5-2.0 keV surface
brightness distribution shows sharp edges to the southwest, southeast
and north, reminiscent of those in the clusters A2142 and A3667. In
addition to the edges, there are two X-ray peaks. The first, to the
east, coincides with the nucleus of NGC507. 
A second peak, $1'$ to the west has no optical
counterpart. However, comparing the X-ray and the radio map 
(Fig.~\ref{n507}a, c) shows that the western radio lobe lies
precisely in the surface brightness trough between the nucleus and the
peak to the west. Thus, it seems likely that the radio lobe, probably
a buoyant bubble, has displaced X-ray emitting gas generating a trough
in the X-ray surface brightness distribution.

The origin of the peculiar sharp surface brightness discontinuities
around NGC507 is unclear. The bright emission is well fit by a thermal
model with gas temperatures near 1 keV, consistent with the mean ASCA
temperature of $1.10\pm0.05$ keV~\cite{matsumoto1997}. The emission
from the central region is resolved and hence the contribution from a
central AGN is relatively small (see Forman et al.~\cite{moriond} for
additional discussion of NGC507). Perhaps the X-ray surface brightness
features arise from motion of NGC507 and its dark halo within
the larger group potential as suggested for the multiple edges in
clusters~\cite{a1795}. 

\section{Conclusions}

We did not expect the rich variety of new structures seen in the
Chandra high angular resolution
observations of clusters and early type galaxies.  Instead of
confirming our prejudices, Chandra has brought us a wealth of new
information on the interaction of radio sources with the hot gas in
both galaxy and cluster atmospheres. We see ``edges'' in many systems
with hot and cold gas in close proximity and have been able to extract
important new parameters of the ICM from their study.  We have only
barely begun to digest the import of the Chandra cluster and galaxy
observations.  We can only expect the unexpected as Chandra
observations continue and as our understanding of how best to use this
new observatory matures.

We acknowledge support from NASA contract NAS8 39073, NASA grants
NAG5-3065 and NAG5-6749 and the Smithsonian Institution.

\end{document}